\documentstyle[prd,aps,epsf,preprint]{revtex}

\def\be{\begin{equation}}
\def\ee{\end{equation}}

\begin{document}
\draft
\author{J. Pantaleone}
\address{ Dept. of Physics, University of Alaska Anchorage }
\author{T.K. Kuo and S.W. Mansour }
\address{Dept. of Physics,  Purdue University }

\title{Constraints on Exotic Mixing of Three Neutrinos}

\maketitle 

\begin{abstract}

Exotic explanations are considered for atmospheric neutrino observations.
Our analysis includes matter effects and the mixing of all three neutrinos under 
the simplifying assumption of only one relevant mixing scale.
Constraints from accelerator, reactor and solar neutrinos are included. 
We find that the proposed mixing mechanisms based on violations of Lorentz invariance or
on violations of the equivalence principle
cannot explain the recent observations of atmospheric neutrino mixing. 
However the data still allow a wide range of energy dependences for the vacuum mixing scale, 
and also allow large electron-neutrino mixing of atmospheric neutrinos.
Next generation long baseline experiments will constrain these possibilities.

\end{abstract}

\section{Introduction}

The SuperKamiokande detector~\cite{SKatmos} has observed a large
deficit in the number of muon-neutrinos generated in the atmosphere by cosmic rays. 
This deficit increases as the propagation distance of the neutrinos increases.
Qualitatively similar results were observed by previous experiments, e.g.~the Kamiokande \cite{Kam}, IMB \cite{IMB} 
and Soudan detectors \cite{Soudan}. The straightforward explanation of the data is that 
the neutrino flavor states are non-degenerate, and so the neutrino flavor changes as the neutrinos propagate.  
This non-degeneracy is readily explained in the Standard Model by small
neutrino masses (Standard Mechanism). 
However other explanations of the atmospheric neutrino observations have been proposed
that do not require neutrino masses. 
One possibility is that the degeneracy of the neutrinos is broken by a Violation of the Equivalence Principle (VEP)
(this was proposed by Gasperini~\cite{Gasperini} and independently by Halprin and Leung~\cite{Halprin}). 
Another possibility is that the neutrino degeneracy is broken by a violation of Lorentz Invariance (LIV)
(this was proposed by Coleman and Glashow \cite{LIV}).  
These two possibilities, VEP and LIV, produce identical neutrino mixing effects \cite{SGV},
which differ from those of the Standard Mechanism.
In this paper we find the constraints imposed on these and other nonstandard neutrino
oscillation mechanisms by accelerator, atmospheric, reactor and solar neutrino observations.

The important difference between the standard mass mixing mechanism and the VEP/LIV scenario is the way the neutrino 
oscillation probability depends on energy.  For either oscillation mechanism,
the conversion probability between two neutrino flavors $\alpha$ and $\beta$ is written as
\be
P(\nu_\alpha \rightarrow \nu_\beta)= \frac{1}{2} \sin^2 2\theta 
[ 1 - \cos \left({2 \pi L\over \lambda }\right) ],
\label{P2}
\ee
where $L$ is the propagation length and $\theta$ is the mixing angle between the two flavors.
This expression neglects the mixing with the third neutrino and also background matter effects. 
For the mass mechanism,
\be
\lambda_{\rm mass} = {4 \pi E \over \Delta m^2} = 
2,500 {\rm km} \left( {10^{-3} eV^2 \over \Delta m^2} \right) \left( {E \over 1 GeV} \right)
\label{mlam}
\ee
where $\Delta m^2=m_2^2-m_1^2$ is the difference in neutrino masses squared and $E$ is the neutrino energy. 
In the case of the VEP scenario, the oscillation wavelength is given by
\be
\lambda_{\rm VEP} = {\pi \over E | \phi | \Delta \gamma } = 
63 {\rm km} \left( {10^{-20} \over |\phi| \Delta \gamma } \right) \left( {1 GeV \over E} \right)
\label{Vlam}
\ee
where $\Delta \gamma = \gamma_2 - \gamma_1$ is the difference in neutrino gravitational couplings
and $|\phi|$ is the magnitude of the gravitational potential.
The potential $|\phi|$ is taken to be constant over the region of interest,  
since $|\phi| = GM/r $ is dominated by distance sources
such as the Great Attractor ($|\phi| \sim 3\times 10^{-5}$) \cite{HLP}. 
In the LIV scenario, the oscillation wavelength is identical to Eq. (\ref{Vlam}) with 
the notational change
\be
|\phi| \Delta \gamma \rightarrow \frac{\Delta v}{2} 
\ee
where $\Delta v = v_2 - v_1 $ is the difference in velocity between the velocity eigenstates. 
Herein we use the VEP notation in our presentation of neutrino data analyses.

The Standard mixing mechanism has strong theoretical motivation.
The VEP/LIV mixing mechanism is less well motivated, 
but it inspires the phenomenological question \cite{atmosVEP}, 
`How well determined is the energy dependence of the neutrino vacuum oscillation wavelength'?
Here we consider, besides the VEP/LIV mixing mechanism, a 
generalized neutrino mixing mechanism with a vacuum oscillation wavelength
\be
\lambda_{\rm gen} = {2 \pi \over \alpha E^n }
\label{Glam}
\ee
where $\alpha$ and $n$ are continuous variables.  The Standard mixing mechanism is $n=-1$ and any other value, 
such as the the VEP/LIV mixing mechanism with $n=+1$, is exotic. 
Here we also analyze neutrino data in this generalized mixing scheme.

The different energy dependences of the mixing mechanisms influence
how neutrino data are analyzed and interpreted.
For example, atmospheric neutrino observations imply large mixing between the muon-neutrino ($\nu_\mu$) 
and the tau-neutrino ($\nu_\tau$),
but are relatively insensitive to electron-neutrino ($\nu_e$) mixing \cite{jimPRL}.
Chooz \cite{Chooz} reactor observations constrain the $\nu_e$ mixing
of atmospheric neutrinos in the Standard Mechanism, but not in the VEP/LIV mechanism.
Thus simplifying assumptions, such as that only two-neutrino oscillations are relevant
for atmospheric neutrinos, are valid only for particular energy dependences of the mixing mechanism.
A similar situation occurs for solar neutrinos.
Solar neutrino observations imply mixing of $\nu_e$.
In the Standard Mechanism the atmospheric and solar observations imply two different oscillation scales,
but in the VEP/LIV mechanism the same scale may suffice for both \cite{HLP}.
Thus it is clearly necessary to include the mixing of all three neutrinos when
attempting to describe the various neutrino observations in an exotic mixing mechanism.

In section II, we describe our three flavor oscillation formalism.
Herein we adopt the simplifying approximation that only one neutrino mixing scale is relevant.
In Section III, we find the constraints on oscillation parameters from accelerator experiments.
Recent results from the CCFR \cite{CCFR} experiment are analyzed in detail to yield new,
more stringent limits on VEP/LIV parameters.
The LSND experimental results are also examined.
In Section IV, a $\chi^2$ analysis is performed on the recent atmospheric neutrino results from the 
SuperKamiokande detector.  
We attempt to fit the contained and throughgoing muon data using exotic neutrino mixing parameters.
In Section V, a $\chi^2$ analysis is performed on the available solar neutrino data.
The accelerator results strongly constrain the 
VEP/LIV interpretations for the solar and atmospheric neutrino observations. 
In Section VI, we summarize our conclusions.

\section{Three-Neutrino Oscillations}

The mixing of three neutrinos can depend on several parameters. 
If the neutrino mixing is caused by any one of the Standard, VEP or LIV mechanisms,
then the observable parameters are two mixing scales, three mixing angles and a phase.
If more than one of these mechanisms occur, 
then the number of observable parameters is more than doubled \cite{HLP}. 
Here we adopt the simplest possible formalism that still allows all three neutrinos to mix.
We make the simplifying approximation that only one mixing scale is relevant.
Then there are only two mixing angles which are observable.
This one-mixing scale approximation allows all three neutrinos to mix---it has  
been used for the Standard Mechanism (see e.g. \cite{jimPRL,1ms} and references therein), 
and here we apply it to exotic mixing mechanisms.

The parameterization of the mixing 
\begin{equation}
| \nu_\alpha > \ \ = \ \ U_{\alpha i} | \nu_i >
\end{equation}
between the flavor eigenstates, \(\alpha = e, \mu, \tau\),
and the energy eigenstates, \(i = 1, 2, 3\), 
is here chosen to be \cite{jimPRL}
\begin{equation}
U = \left[ \begin{array}{ccc}
0 &  \cos \phi & \sin \phi \\
- \cos \psi & - \sin \psi \sin \phi & \sin \psi \cos \phi \\
  \sin \psi & - \cos \psi \sin \phi & \cos \psi \cos \phi 
\end{array} \right]
\label{U}
\end{equation}
where \(\phi\) and \(\psi\) are the mixing angles
(N.B. We adopt here the notation of previous papers, 
but warn the reader not to confuse 
the $\nu_e$ mixing angle $\phi$ with
the gravitational potential $|\phi|$.)
This parameterization is chosen 
such that matter effects \cite{W} are straightforward. 
In a matter background, \(\psi\) is unchanged but the
effective \(\phi\) is given by
\begin{equation}
\sin^2 2 \phi_m = { { ( S \sin 2 \phi )^2 } \over
{ (T - S \cos 2 \phi )^2 + ( S \sin 2 \phi )^2 } }  \  \ .
\end{equation}
and the potential eigenvalues are 
\begin{eqnarray}
V_1 &=& 0 \label{V} \\
V_{3,2} &=& {1 \over 2} [ ( S + T ) \pm 
\sqrt{ (T - S \cos 2 \phi )^2 + ( S \sin 2 \phi )^2 } ]
\nonumber
\end{eqnarray}
where the $i = 2$ state is associated with the minus sign. 
Here \(T\) is the neutrino potential induced by forward scattering
off of the electron background, 
\begin{eqnarray} 
T = \sqrt{2} G_F N_e  
\label{T}
\end{eqnarray}
with \(G_F\) as Fermi's constant, and $N_e$ the number density of electrons.
For antineutrinos, \(T \rightarrow - T\).  
The neutrino potential $S$ is the vacuum contribution
\be
S = \left\{ 
\begin{array}{l l}
   {\Delta m^2 \over 2 E} & \mbox{Standard mass mixing} \\
  2 |\phi| \Delta \gamma E & \mbox{VEP/LIV mixing} \\
  \alpha E^n               & \mbox{Generalized  mixing}
\end{array}
\right.
\label{S}
\ee
The generalized mixing is introduced to allow us to probe a wide class
of exotic mixing scenario's without regard to theoretical prejudice.
Generalized mixing includes the Standard mixing mechanism, $n=-1$,
and the VEP/LIV mixing mechanism, $n=+1$.

To illustrate the physical implications of this parametrization,
we give the relevant oscillations probabilities
for a constant matter density. 
\begin{eqnarray}
P(\nu_e \rightarrow \nu_e) &=&  
1 - {1 \over 2} \sin^2 2 \phi_m [1 - \cos ( \beta_3 - \beta_2 ) ]
\nonumber \\
P(\nu_\mu \rightarrow \nu_e) &=&
{1 \over 2} \sin^2 \psi \sin^2 2 \phi_m [1 - \cos ( \beta_3 - \beta_2 ) ]
\label{PPP}
\\
P(\nu_\mu \rightarrow \nu_\mu) &=&
1  - {1 \over 2} \{ \sin^2 2 \psi [ 1 
- \sin^2 \phi_m \cos (\beta_2 - \beta_1)
- \cos^2 \phi_m \cos(\beta_3 - \beta_1) ] \nonumber \\
& & + \sin^4 \psi \sin^2 2 \phi_m [ 1 - \cos ( \beta_3 - \beta_2 ) ] \}
\nonumber 
\end{eqnarray}
Here the dynamical phase acquired by a neutrino energy eigenstate which
propagates for a time $t$ is 
\begin{equation}
\beta_i \equiv V_i t 
\end{equation}
Unitarity and time reversal symmetry 
can be used to obtain all other oscillation
probabilities from those above.

The above equations describe a constant density medium. 
For more complicated density distributions, 
the procedure for calculating the probabilities is analogous
to those used for massive neutrinos (see e.g. \cite{KP}).
Note that the length scale associated with matter effects is
\be
\lambda_{\rm matter} = {2 \pi \over T} = 6,600 {\rm km} 
\left( \frac{2.5 {\rm g / cm^3}}{Y_e \rho}  \right)
\label{lammat}
\ee
where $Y_e$ is the number of electrons per nucleon, $\rho$ is the density,
and we have chosen a value corresponding to roughly the average for the Earth's mantle.
Because this length is of order the radius of the Earth, 
matter effects must be included in three-flavor analyses of atmospheric neutrinos \cite{jimPRL,1ms}.

In the two-flavor {\it vacuum} approximation, 
neutrino oscillation effects are symmetric for mixing angles 
in the ranges 0 to \(\pi/4\) and \(\pi/4\) to \(\pi/2\).
However when there are three flavors (and also when matter effects
are relevant) there is no symmetry between these two ranges.
Thus we use limits where  \(\phi \) and \(\psi\) explicitly range 
between 0 and \(\pi/2\), or, equivalently, the sines of these angles
range between 0 and 1.  In addition, the mixing scale parameter in $S$
may be positive or negative \cite{1ms},
but this sign is only observable when matter effects are important.
This covers the full allowed range for these parameters,
without any redundancy.

When one of the mixing angles is at the limit of its range,
then one of the neutrino flavors decouples and 
the approximate three-flavor notation 
(Eq. (\ref{U})) reduces to a two-flavor 
description in terms of the remaining mixing angle (see Table (1)).  
All possible two-flavor approximations are included.
However since there are  only two mixing angles, 
the three possible types
of two-flavor approximation are not fully independent.
This just follows from the assumption of
only one relevant mass scale.

\section{Accelerator Experiments}

Solar and atmospheric neutrino observations probe small neutrino mixing scales because
they involve long neutrino propagation distances. 
Accelerator experiments typically have much smaller baselines and higher energies.
The higher energies are advantageous for probing some exotic neutrino mixing mechanisms ($n > 0$).
In particular, the sensitivity of an experiment to the VEP/LIV mechanism is proportional to $L E$, see Eq. (\ref{Vlam}).  
Thus the higher energies of accelerator experiments compensates for their shorter baselines.
Accelerator experiments provide a limit on the VEP/LIV mixing scale which is
comparable to those coming from solar and atmospheric neutrino observations.

Accelerator experiments have been analyzed previously for their constraints on the VEP/LIV mechanism \cite{HLP}.
The limits on the breaking scale from accelerator experiments have drastically improved recently with
the results from the CCFR experiment~\cite{CCFR} at Fermilab. 
Here we describe our detailed analysis of this experiment's results.  
We obtain two-flavor constraints for specific channels, 
and then derive a useful three-flavor constraint.
Specific implications of these constraints for the interpretations of the solar and 
atmospheric neutrino observations will be discussed in later sections.
Finally we briefly discuss the LSND \cite{LSND} results.

The CCFR neutrino beam consists of roughly 70\% $\nu_\mu$, 28\% ${\bar \nu}_\mu$, 1.6\% $\nu_e$ and 0.6\% ${\bar \nu}_e$.
The experiment has a 1.4 km oscillation length of which 0.5 km is the decay region.
Our analysis is based on the published electron spectrum \cite{CCFR} as a function of visible energy. 
There are 15 energy bins running from 30 GeV to 600 GeV. 
The $\chi^2$ function used in the analysis is given by
$$\chi^2=\sum_{i=1}^{15} {\left(N^{\rm data}_i-N^{\rm th}_i\right)^2\over \sigma_i^2},$$
where $N^{\rm data}_i$ and $\sigma_i$ are the binned experimental values and errors, respectively. 
$N^{\rm th}_i$ is the theoretical prediction of the electron flux for the energy bin $i$ assuming oscillations.

The effects of neutrino mixing on the observed $\nu_e$ flux is calculated as
\be
N^{\rm th} = N^{\rm exp} [P(\nu_e\rightarrow\nu_e)+P(\nu_e\rightarrow\nu_\tau) K(E)]
+ r(E) [P(\nu_\mu\rightarrow\nu_e)+P(\nu_\mu\rightarrow\nu_\tau) K(E)] 
\ee
Here $N^{\rm exp}$ is the expected flux of $\nu_e$  calculated without oscillations,
$ r(E) \sim 100$ is the ratio between the produced $\nu_\mu$ and $\nu_e$ fluxes, 
and $K(E)$ is the probability of misidentifying a $\nu_\tau$ as a $\nu_e$ (18\%) times
the ratio of the $\nu_\tau$ to the $\nu_e$ nucleon cross-section~\cite{PrivateC1}.
In our analysis, we have taken the neutrino energy to be equal to the visible energy in the detector.
The smearing of the neutrino oscillation phase has been accounted for by 
averaging the neutrino propagation length over the length of the decay pipe and 
by averaging the neutrino energy over the bin width.  
Limits on the mixing scale are most sensitive to the high (low) energy end of 
the electron spectrum in the VEP/LIV (Standard) mixing mechanism.
To be conservative in our analysis, we inflated the error bar on the highest energy data point by a factor of two.

The oscillation probabilities are as described in section II,
with the simplification that matter effects are negligible because the
accelerator baseline is so much shorter than the matter length scale, Eq. (\ref{lammat}).
The two-flavor oscillation limits are calculated by taking the limiting values
described in Table (1).  For example, for two-flavor $\nu_\mu\rightarrow\nu_e$ oscillations
$P(\nu_e\rightarrow\nu_\tau) = P(\nu_\mu \rightarrow \nu_\tau) = 0$, 
$P(\nu_e\rightarrow\nu_e) = 1- P(\nu_\mu \rightarrow \nu_e)$ and the latter oscillation probability 
is given by Eq. (\ref{P2}). Because $r(E)$ is larger than 1, 
the net effect of these oscillations will be an energy-dependent increase in the observed $\nu_e$ flux.
Similarly, for two-flavor $\nu_e\rightarrow\nu_\tau$ oscillations 
$P(\nu_e\rightarrow\nu_\mu) = P(\nu_\mu \rightarrow \nu_\tau) = 0$, 
$P(\nu_e\rightarrow\nu_e) = 1- P(\nu_e \rightarrow \nu_\tau)$ and,
because $K(E)$ is smaller than 1, 
these oscillations will decrease the observed $\nu_e$ flux.
Finally, for two-flavor $\nu_\mu \rightarrow\nu_\tau$ oscillations,
$P(\nu_e\rightarrow\nu_\mu) = P(\nu_e \rightarrow \nu_\tau) = 0$,
$P(\nu_e\rightarrow\nu_e) = 1$ and,
because $r(E) K(E)$ is greater than 1,
these oscillations will increase the observed $\nu_e$ flux.

Fig.~(1) shows the 90\% exclusion region for the two-flavor $\nu_\mu\rightarrow\nu_e$,  
$\nu_e\rightarrow\nu_\tau$ and $\nu_\mu\rightarrow\nu_\tau$ oscillations assuming the VEP/LIV scenario. 
The $\chi^2$ value for the contours are chosen such that the limits on the mixing angles at 
large mixing scale agree with those found for the Standard mechanism \cite{CCFR}.
The limiting values of our two-flavor results are summarized in Table. (2).
The limits on the VEP/LIV mixing scale from the $\nu_\mu$-$\nu_e$ oscillation parameters are much more severe
than those from $\nu_e\rightarrow\nu_\tau$ oscillations, because the former can produce
a 50-fold enhancement in the $\nu_e$ flux while the latter reduces the flux by at most 50\%.
The limits on $\nu_\mu\rightarrow\nu_\tau$ oscillations are intermediate between the previous two.

For three-flavor mixing, the two-flavor constraints in Table (2) are no longer valid.
With more than one oscillation channel acting the effects of mixing may 
enhance or suppress the $\nu_e$ flux, depending on the mixing parameters.
As an example, we give here a simple, approximate expression for a
constraint on the three-flavor mixing parameters in the VEP/LIV scheme.
Using the vacuum oscillation probabilities 
and making the small phase approximation,
the 90\% confidence level contour is approximately given by
\begin{eqnarray}
1 = \left( \frac{|\phi| \Delta \gamma}{10^{-23}} \right)^2  \cos^2 \phi 
|   &\sin^2&  \phi [-A + B   \sin^2 \psi + C  \cos^2 \psi ] 
 \nonumber \\
   &+&  D  \sin^2 \psi \cos^2 \psi \cos^2 \phi |
\label{acc3}
\end{eqnarray}
where $A, B, C, D$ are constants which we determine to be
\begin{eqnarray}
        A & = & 9 \times 10^{-3} \nonumber \\
        B & = & 9 \times 10^{-1} \nonumber \\
        C & = & 1.6 \times 10^{-3} \nonumber \\
        D & = & 1.6 \times 10^{-1} \nonumber 
\end{eqnarray}
This expression is related to the two-flavor, maximal mixing limits given in Table (2).
It is only valid when the net mixing amount is `large'.
We note that a cancellation in the net mixing occurs for $\sin^2 \psi \leq 0.01$.
However since atmospheric neutrino observations suggest large $\nu_\mu - \nu_\tau$ mixings of
($\psi \sim \pi/4$), this cancellation is not relevant.
We shall use this expression for comparisons with atmospheric and solar neutrino results.

The LSND experiment has found evidence for neutrino mixing \cite{LSND}.
The data indicate mixing at a very low level, P($\nu_\mu \rightarrow \nu_e$) = $(0.31 \pm 0.12) \% $. 
The main evidence for neutrino mixing comes from observations of muon decay at rest (DAR),
however there is also a weak indication of mixing from observation of pion decay in flight (DIF).
The LSND result has not been confirmed by independent experimental groups.
Here we use the framework of generalized neutrino mixing, Eq. (\ref{S}),
to examine the LSND results and how they collate with other experimental results.

The LSND experiment has a propagation length of $30 \pm 4$m. The DAR data 
involves neutrino energies in the range $20-60$ MeV, while the DIF
data involves neutrino energies of $60-200$ MeV.  The DAR data
spans a relatively small range of energies and is not sensitive to the energy 
dependence of the neutrino mixing (see Fig. (31) of the Phys.~Rev.~C article in \cite{LSND}).  
The DAR and the DIF
data sets involve very different energy scales, so a comparison of the two sets
can probe the energy dependence of the vacuum mixing---{\it if the mixing angle is large}.  
For the allowed regions at small mixing angle, there is generally no sensitivity to the mixing
scale because the oscillations average out.

The CCFR experimental results exclude two-flavor, $\nu_\mu - \nu_e$ oscillations
for $\sin^2 2 \theta > 1.8 \times 10^{-3}$ for large mixing scales. This is
a small enough mixing angle to conflict completely with the LSND results if the mixing scales overlap.
The relative parameter regions probed by the CCFR and LSND experiments in the generalized
exotic neutrino mixing framework can be determined by
comparing the vacuum oscillation phase of the CCFR experiment with that of the LSND experiment
\be
{S_{\rm L} L_{\rm L} \over S_{\rm C} L_{\rm C} }
=
{\alpha_L \over \alpha_C} {{L}_{\rm L} \over {L}_{\rm C} } \left({{E}_{\rm L} \over {E}_{\rm C} } \right)^n
\approx
{\alpha_L \over \alpha_C} 10^{-3.5(n+.45)}
\label{LCphases}
\ee
where the subscripts L and C denote the LSND and CCFR quantities, respectively,
and Eq. (\ref{S}) has been used for S.
For the LSND experiment, we have used that the neutrinos are characterized by an average energy
of 40 MeV and a propagation length of 30 m.  
The oscillation phase determines the threshold of observability of oscillations,
as the phase increases from a small value to a value of order 1, the observability of oscillations increases.
Eq. (\ref{LCphases}) shows that for a common mixing scale, $\alpha_L = \alpha_C$,
the mixing effects observed by LSND would not be in conflict with the CCFR results only for
\be
n < -0.45 
\label{nacc1}
\ee
This agrees with previous observations
that the results do not conflict for the Standard mixing mechanism, $n=-1$,
but do conflict for the VEP/LIV mechanism \cite{noLSND} \cite{HLP}, $n=1$.

Relevant limits on two-flavor, $\nu_\mu - \nu_e$ oscillations also come from the BNL-776 experiment \cite{BNL776}.
This experiment has an average energy of 2.5 GeV and a neutrino propagation length of 1 km.  At large mixing
scales the BNL-776 experiment excludes $\sin^2 2 \theta > 3 \times 10^{-3}$. 
This is enough to conflict with the favored LSND parameters if the mixing scales overlap 
(except for possibly a very small parameter region at small mixing angles which we shall neglect).
Comparing the oscillation phases in the generalized framework 
it follows that the LSND and BNL-776 results would not overlap only for
\be
n < -0.85
\label{nacc2}
\ee
This agrees with the observations that the results do not conflict for the Standard mixing mechanism, $n=-1$.

Reactor experiments are $\nu_e$ disappearance experiments, 
and they have set limits on mixing with $\nu_e$ at relatively large values of the mixing angle.
However, unlike accelerator experiments,
reactor experiments have a smaller average energy than the LSND experiment so they are sensitive to
a different range of $n$.
The Chooz reactor experiment \cite{Chooz} has an average energy of 3 MeV, an average propagation length of 1 km,
and excludes $\sin^2 2 \theta > 0.18$ at large mixing scales.  
The large mixing solutions of the LSND data do not conflict with the CHOOZ results for
\be
1.4 < n
\label{reacn}
\ee
Comparing this with Eqs. (\ref{nacc1}) and (\ref{nacc2}), we see that the LSND 
favored region at large mixing angles is in conflict with
reactor and/or accelerator experiments for any $n$ value.
Therefore solutions to the LSND data are restricted to small $\nu_e$ mixing angles.

A comparison of LSND and atmospheric vacuum oscillation phases suggests that
at least two different mixing scales, $\alpha_{\rm LSND} \neq \alpha_{\rm atmos}$,
are required.  This is because the $n$ value when the scales coincide, 
$n \approx -3$, is not allowed by the atmospheric neutrino data (see Section IV).
However this leaves no room for an explanation of the solar neutrino observations.
For three neutrinos there are only two independent energy splittings,
and the best explanations of the solar neutrino results require another mixing scale (see Section V). 
Marginal explanations of the solar data are possible without another mixing scale,
if one has large $\nu_e$ vacuum mixing.
But reactor experiments restrict $\nu_e$ mixing to be small at the LSND scale
and at the atmospheric scale for the LSND allowed $n$ range given in 
Eq. (\ref{nacc2}) (see Section IV).
Thus the situation is similar to the situation for the Standard mixing mechanism, $n=-1$,
where an acceptable, three neutrino (two mixing scale) explanation for
the LSND, atmospheric and solar data does not exist (see e.g. \cite{FogliLE} and references therein).
Generalized neutrino mixing does not enable a good fit to all of the neutrino mixing data.

The miniBooNE experiment \cite{boone} is under construction at Fermilab.  
It will have a propagation length of $L \sim 500$ m and an energy range of
$0.1 < E < 1$ GeV.  It will probe the LSND result unless $ n < -1.1 $.
Thus this experiment will test the Standard mixing mechanism.

\section{Atmospheric Neutrino Experiments}

The large SuperKamiokande neutrino detector has observed many more atmospheric neutrinos than ever before
and has confirmed the suggestive effects observed by previous detectors Kamiokande \cite{Kam}, 
IMB \cite{IMB} and Soudan \cite{Soudan}. 
Atmospheric neutrinos have been measured with energies ranging from 0.1 to 1000 Gev, 
and with propagation distances ranging from 10 to 10,000 kilometers.
In general, the SuperKamiokande detector observes the expected $\nu_e$ flux however the 
$\nu_\mu$ flux is much less than expected, 
and the size of this deficit increases with propagation distance.  
It is clear evidence for new neutrino physics.
Here we attempt to explain the data with exotic neutrino mixing mechanisms.

The most recently published contained, partially contained and throughgoing SuperKamiokande data 
\cite{SKatmos} are used in our analysis.  
We follow the analysis method used by the SuperKamiokande group to perform our oscillation fit.
A $\chi^2$ quantity is constructed
\begin{equation}
\chi^2 = \sum_{p, \cos \theta, e, \mu} {\left(N_{\rm data}-N_{\rm th} \right)^2 \over \sigma^2} 
+ \sum_{j} {\epsilon_j^2 \over \sigma_j^2 }
\end{equation}
Here $N_{\rm data}$ is the number of data events in the bin as given in Ref. \cite{SKatmos}, 
$\sigma$ is the statistical error on the data and Monte Carlo,
$N_{\rm th}$ is the theoretically predicted number of events, $\sigma_j$ is a systematic error and  $\epsilon_j$ is varied
to minimize the $\chi^2$.
The contained and partially contained data is broken up into 2 flavors $\times$ 5 zenith angle bins $\times$
4 energy bins to yield 40 separate bins, and the upward, throughgoing $\mu$ data is in 10 zenith angle 
bins---thus we work with a total of 50 data bins.  
Particle misidentification has been included according to the SuperKamiokande Monte Carlo estimates.  
For the systematic errors, 
we only include in our analysis the largest two of those listed in Ref. \cite{SKatmos}, 
the flux normalizations of the contained/partially contained data and that of the upward, throughgoing muon data.

The theoretically expected flux values are calculated in the one mixing scale
framework described in Section II.  Analytical expressions for the oscillations
probabilities through the earth are used, with the density distribution of matter in the
Earth approximated by two densities, one for the core and one for the mantle.   
Since the size of these structures is smaller than the length scale associated with matter
effects (see Eq. (\ref{lammat})) this is a reasonable approximation.
The energy and angle dependence of the atmospheric neutrino flavor ratio and of the neutrino-antineutrino
ratio are taken from Ref. \cite{Honda}.  
The energy distribution for a particular bin is taken from various references \cite{SKatmos} \cite{other}.
The angular distributions of neutrinos for a particular zenith angle bin is calculated using
the energy dependent average scattering angle of SuperKamiokande \cite{angle}.
The distribution of neutrino production points is taken from Ref. \cite{pathlength}.
The oscillations probabilities are averaged over these distributions, then multiplied by the 
expected flux without oscillations \cite{SKatmos} to get $N_{\rm th}$.

Assuming the VEP/LIV mechanism is the only source of neutrino mixing,
we have attempted to fit the SuperKamiokande data. 
The quality of the fit is indicated in Fig. (2) where 
chi-squared contours are shown on a plot of the mixing scale $|\phi| \Delta \gamma$ versus the mixing parameter
$\sin^2 \phi$ which describes the amount of mixing with the $\nu_e$.
The solid line surrounds the 
parameter region allowed by the contained and partially contained data at 95\% C.L. 
($\chi^2 =52$ for $40-2$ plot parameters $-1$ floating parameter $= 37$ degrees of freedom)
and the dashed line surrounds the
parameter region allowed by the throughgoing data at 95\% C.L.  
($\chi^2 = 14$ for $10 - 3 = 7$ degrees of freedom).
The value of $\psi$  at each point in this plot is varied to minimize $\chi^2$.
Note that these two data sets were analyzed 
{\it separately} so the values of $\psi$ in the two analyses 
was generally rather different at each point on the graph. 
We also did a combined analysis, with a common $\psi$ value, 
and found $\chi^2_{\rm min} = 87.5$ for 50-3=47 degrees of freedom. 
This is obviously a very bad fit to the data, as it
corresponds to a probability of being the correct interpretation of order $3\times10^{-4}$. 
A filled circle has been place on Fig. (2) to indicate the location of the minimum.
In general, we did not find an acceptable fit to all of the atmospheric neutrino data for the VEP/LIV mechanism.

The nature of the `separately allowed' regions in Fig. (2) and the lack of a common fit can 
be easily understood.  
Previous analyses \cite{atmosVEP} have found that there is no allowed two-flavor, $\nu_\mu-\nu_\tau$ parameter region,
and Fig. (2) confirms that since there is no allowed region at the left edge of the graph where $\sin^2 \phi = 0$ 
(see Table (1)). 
The two `separately allowed' parameter regions are at large $\nu_e$ mixing
where matter effects are substantial, and at relatively large values of
the mixing scale where vacuum oscillations are suppressed.
In this region, the effect of neutrino mixing is given approximately by
\begin{eqnarray}
{ N_\mu \over N_\mu^0} & \approx & \left[ 1 - \sin^2 \psi + 2 \sin^4 \psi \right ]
- {1 \over 4} \sin^2 2 \psi \ \left[ 1 - 
\cos \left( \pi t \over \lambda_{matter}  \right)  \right]
+ {1 \over 2 R } \sin^2 \psi
\\
{ N_e \over N_e^0} & \approx & {1 \over 2} \left[ 1 + R \sin^2 \psi \right]
\end{eqnarray}
Here $N_\mu^0$ and $N_\mu$ ($N_e^0$ and $N_e$) are the muon (electron) fluxes that are expected without oscillations
and what is actually observed with oscillations, respectively,
and $R \equiv {N_\mu^0 \over N_e^0} \approx 2$ is the ratio of $\nu_\mu$ to $\nu_e$ expected without oscillations.
This approximate expression has been derived using the analytical oscillations expressions
of Eq. (\ref{PPP}), taking $\cos^2 \phi \approx 0.5$ (as indicated by Fig. (2)), 
and taking the average of oscillation terms involving the large vacuum phase to vanish.
The electron signal is independent of angle and is approximately the no oscillation flux
for $\sin^2 \psi \approx 0.5$, while the muon signal will show an angular dependence
from matter induced oscillations.  These energy independent oscillations occur
because matter effects split the degeneracy between lowest two energy eigenstates 
(see \cite{jimPRL} and \cite{1ms} and references therein).
The amplitude of the matter oscillations, and hence the zenith angle variations,
depends on $\sin^2 2 \psi$ where $\psi$ is the value of the $\nu_\mu -\nu_\tau$ mixing angle.
For the contained/partially contained data the zenith angle variations are very large,
of order 50\%, so the data require $\sin^2 2 \psi \approx 1 $.
For the upward, throughgoing muon data the zenith angle variations are rather small,
of order 25\%, and the data require $\sin^2 2 \psi \approx 0.5 $.  
In the Standard mixing mechanism this amplitude difference is explained because energy dependent 
smearing of the vacuum oscillations reduces the size of oscillation effects more for
the throughgoing muon data,  but for energy independent matter oscillation this smearing is absent.
An acceptable solution to both data sets can not be found
for a common value of $\psi$ in the VEP/LIV mixing mechanism.

Also shown on Fig. (2) are the implications of the accelerator neutrino experiments.
The dotted contour is generated using Eq. (\ref{acc3}), where the values of $\sin^2 \psi$
are taken from the best fit to the total data.
The region above the dotted line is excluded by the accelerator data.
The `separately allowed' and the (unacceptable) best fit to the total data are well inside the
excluded region.  Below the accelerator limit, the smallest chi-squared value for
the fit to all of the atmospheric neutrino data is $\chi^2=107$ for 50-3=47 degrees of freedom,
which corresponds to a $1.4 \times 10^{-6}$ probability of being the correct interpretation.
Thus the accelerator data make the VEP/LIV interpretations of the atmospheric 
neutrino data even less tenable.

The results of using the generalized neutrino mixing framework to describe
atmospheric neutrinos are shown in Fig. (3).  
Here is plotted the minimum value of chi-squared for the combined contained, partially contained 
and the upward, throughgoing muon data (50 bins) as a function of the energy exponent $n$.
The length $L$ and energy $E$ dependence of the vacuum oscillation phase for various $n$ values
are explicitly labeled on the graph.
The mixing scale parameter $\alpha$ and the mixing angles $\psi$ and $\phi$ have been varied
so as to minimize chi-squared at each $n$ value for the dotted contour.
The dashed line shows the analogous two-flavor, $\nu_\mu-\nu_\tau$ mixing results ($\phi = 0$).
The solid, horizontal line indicates the  99\% C.L. ($\chi^2 = 71$ for $50-4=46$ degrees of freedom).

The two-flavor contour on Fig. (3) (dashed line) agrees with that of Fogli et. al. \cite{atmosVEP}.
In the two-flavor limit the data clearly prefer the Standard mixing mechanism ($n=-1$),
with only a small region around that value allowed.
However for the more general three-flavor case, the added freedom of allowing
mixing with the $\nu_e$ substantially changes the situation.  Then the atmospheric neutrino data
allow a wider range of energy dependences, approximately $n < 0$.

Also on Fig. (3) are the constraints from accelerator (Eq. (\ref{acc3})) 
and reactor \cite{Chooz} neutrino experiments.  In the atmospheric neutrino data, 
oscillation effects show up at distances as small as hundreds of kilometers. 
For $n > 0$ ($n < 0$), the oscillation wavelength
decreases as energy increases  (decreases) so experiments with higher (lower) energies than atmospheric neutrinos
such as accelerator (reactor) neutrino experiments can probe the observed neutrino mixing at shorter distances.
Fig. (3) shows that accelerator experiments exclude the best fit values for $n > 0.9$
which includes the VEP/LIV mixing mechanism at $n=1$.  The precise $n$ value of the accelerator limit
is relatively insensitive to whether the mixing is two-flavor or three-flavor.
Reactor experiments are only sensitive to $\nu_e$ mixing, so they only exclude 
the best fit three-neutrino constraint for $n < -0.8$.
This agree with the general observation for the Standard mixing mechanism, $n=-1$, 
that the Chooz reactor neutrino measurements limit the size of the $\nu_e$ mixing 
in the region allowed by the atmospheric neutrino data.
However note that we have not explored how much the allowed $n$ region can be expanded
by some small amount of $\nu_e$ mixing compatible with the reactor experiments.

Fig. (3) indicates that the currently allowed range of $n$ values is roughly $-1.6 < n < 0$.
Large $\nu_e$ mixing of the atmospheric neutrinos is possible for roughly $-0.8 < n < 0$.
Solar neutrinos would be affected by this large mixing, so it will be indirectly probed by
future solar neutrino experiments \cite{newsolar}.
Also, next generation long baseline reactor neutrino experiments such as KamLAND \cite{KamLAND}
should be able to extend the reactor limits on Fig. (3) to the right.
The full allowed range of $n$ will be directly probed by next generation long baseline accelerator experiments 
such as the MINOS \cite{MINOS} experiment, which is currently under construction,
and the planned NGS \cite{NGS}.  These experiments will increase the neutrino propagation length
of accelerator experiments by almost three orders of magnitude over previous experiments 
and should be able to move the accelerator limit on Fig. (3) to the left
enough to reach the Standard Mixing mechanism at $n=-1$.

Both plots Fig. (2) and Fig. (3) are for a positive mixing scale ($|\phi| \Delta \gamma > 0$ in the VEP/LIV scenario).  
We have also recalculated all plots for the opposite sign choice ($|\phi| \Delta \gamma < 0$) and have found no
significant difference.  Matter effects make oscillations effects dependent on
the sign of the mixing scale, however as noted elsewhere \cite{jimPRL}, the current atmospheric
neutrino data happen to be relatively insensitive to this sign.

Although the VEP/LIV mixing mechanism does not fit the atmospheric neutrino data,
we have {\it not} attempted to constrain the VEP/LIV parameters from this data.
The atmospheric data clearly indicate mixing, and this must be accounted for.
It is necessary to go beyond the one mixing scale formalism and add
additional parameters which have the capability of explaining the data.
The choice of exactly what parameters to add is not completely obvious.
The Standard mechanism is supported by theoretical prejudice,
but Fig. (3) shows that the data still allow a wide choice of exotic solutions.  
A two mixing scale analysis, involving the Standard mixing mechanism and
an exotic mixing mechanism, must necessarily involve at least 7 parameters and is
beyond the scope of this paper.

\section{Solar Neutrino Experiments}

In the framework of exotic neutrino mixing, 
it is instructive to compare solar neutrino oscillations with those of atmospheric neutrinos.
Neutrino oscillation dynamics are characterized by the oscillation phase,
the ratio of phases is
\be
{S_{\rm a} L_{\rm a} \over S_{\rm s} L_{\rm s} }
=
{\alpha_a \over \alpha_s} {{L}_{\rm a} \over {L}_{\rm s} } \left({{E}_{\rm a} \over {E}_{\rm s} } \right)^n
\approx
{\alpha_a \over \alpha_s} 10^{-2(1-n)}
\label{asphases}
\ee
where the subscripts $s$ and $a$ denote solar and atmospheric neutrino quantities, respectively.
Here we have used that atmospheric neutrinos are characterized by energies of order 1 GeV and lengths
of order the distance to the horizon, which is about 10\% of the Earth's radius.
For solar neutrinos, we have used that they are characterized by energies of order 10 MeV,
and for a solar length scale we have used the shortest possible length scale for which
large, energy dependent flux reductions are possible---the scale of density variations in the Sun, 
which is about 10\% of the Sun's radius.

For $n < 1$ (which includes the Standard mechanism at $n=-1$) 
the indications of mixing imply $\alpha_a >> \alpha_s$ and so the
solar and atmospheric neutrinos are described by different scales.
This conclusion is only enhanced if the larger length scales possible in the Sun are used,
such at the Earth-Sun distance.
Thus describing solar and atmospheric neutrino oscillations simultaneously in a one mixing scale 
framework is generally not possible for $n < 1$.

For the VEP/LIV mixing mechanism, $n=1$, Eq. (\ref{asphases}) shows the previous observation
\cite{HLP} that solar and atmospheric
neutrino oscillation dynamics imply a common mixing scale, $\alpha_s = \alpha_a$.
As we shall show, for $n \geq 1$, accelerator experiments constrain solar neutrinos.

First we give a short discussion on the VEP/LIV solutions from the 
study of the solar neutrino spectrum. 
The reader is referred to Refs.~\cite{HLP,BahcallVEP,MK} for details. 
The latest results from SuperKamiokande~\cite{SKsolar} and 
previous experiments~\cite{Solarprevious} observe a deficit in the 
measured solar neutrino flux when compared to the predictions of the 
Standard Solar Model (SSM)~\cite{Bahcallsolar}. 
Apart from the usual Standard mixing mechanism, 
the VEP/LIV mixing mechanism can explain the observed total rate of neutrino events 
in the different solar neutrino detectors. 
In~\cite{MK}, the analysis using the first 504 days total rate measured 
by the SuperKamiokande gives two allowed solutions: 
the small angle solution at 
($\sin^2 2 \theta\sim 10^{-3}, \left|\phi\right| \Delta \gamma \sim 10^{-18}$), 
and the large angle solution at 
($\sin^2 2 \theta\sim 1, \left|\phi\right| \Delta \gamma \sim 10^{-21}$). 

In this paper, we have updated our results using the first 708 days data~\cite{708} 
run of the SK experiment for the total rate. 
The rate analysis is necessarily unchanged from the one in~\cite{MK} 
as no considerable change in the measurement of the SK total rate 
has occured. 
The main change is in the spectral shape analysis as the 
highest data bin is now considerably lower than the 504 days result.
This will change the spectral shape analysis exclusion region of Ref.~\cite{MK}; however, 
we will not show the spectral shape analysis here due to the current uncertainty of the
hep contribution to the solar neutrino flux and the high energy bins of the recoil electron spectrum. 
In any case, the accelerator bounds do not allow for the solutions from the rate analysis.
In Fig.~\ref{fig;solar}, we show the allowed regions from the rate analysis, together
with the accelerator three-flavor exclusion region at the 90\% CL. 
To calculate the accelerator bound, we have used Eq. (\ref{acc3}) with $\psi = \pi/4$,
as indicated by the atmospheric neutrino data.
It is clear that the CCFR bounds on
the VEP/LIV mechanism exclude both solutions; hence, the VEP/LIV mechanism
is no longer a valid explanation for the solar neutrino data.

\section{Conclusions}

We have analyzed neutrino oscillation data from accelerator, atmospheric, reactor and solar observations  
for evidence of exotic neutrino mixing mechanisms.
Our analysis includes matter effects and the mixing of all three neutrinos, 
albeit in the simplifying ``one mixing scale'' approximation.

In the VEP/LIV mixing scenario, solar and atmospheric neutrino mixing 
are produced at a common mixing scale.   Accelerator experiments have now probed that scale.
The CCFR group has recently announced results on accelerator neutrinos observations at Fermilab.
They found no evidence for neutrino mixing.  
Because their average neutrino energy times neutrino propagation length is larger than that of 
other current terrestrial neutrino experiments,
strong constraints can be placed on VEP/LIV mixing parameters.
We have used the CCFR results to derive constraints on relativity violations down to levels of order
$ | \phi | \Delta \gamma \approx 10^{-22}$ (see Fig. (1), Table (2) and Eq. (\ref{acc3})).

The SuperKamiokande neutrino detector has observed conclusive evidence for neutrino mixing.  
We have attempted to fit the observations using the VEP/LIV mixing mechanism. 
There are acceptable fits to the contained and throughgoing
neutrino data, separately, in a three-neutrino formalism (see Fig.~(2)).
The fit to the combined SuperKamiokande data is not acceptable.
This very poor fit is rendered even more untenable
because it lies inside the parameter region excluded by the recent CCFR accelerator data.
The VEP and LIV exotic mixing scenarios cannot explain the evidence for 
neutrino mixing observed by SuperKamiokande.

Solar neutrino observations also indicate neutrino mixing. 
These observations can be explained by the VEP/LIV mixing mechanism.
However the parameter regions favored by solar neutrino observations lie 
inside the region excluded by the CCFR accelerator neutrino data (See Fig. (4)).
Thus the VEP/LIV mechanism cannot explain the solar neutrino observations.

More exotic explanations of the neutrino data have also been considered.
To remove `theoretical prejudice'
we have let the energy dependence of the one mixing scale vary continuously, 
$\lambda_{\rm gen} = {2 \pi \over \alpha E^n }$.
The suggestions of neutrino mixing from the LSND experiment have been considered in this 
generalized neutrino mixing framework.  
We find that the LSND observations are in direct conflict with constraints
from other accelerator experiments unless $n < -0.85$.  
Adding the atmospheric, solar and reactor results, 
there does not exist a good fit at any $n$ value.

The SuperKamiokande data has been analyzed in this generalized neutrino mixing framework. 
Fig. (3) shows the best fit chi-squared to the combined SuperKamiokande
data as a function of the energy exponent of the mixing scale, $n$.
Energy exponents in the range $-1.6 < n < 0$ are allowed and
large $\nu_e$ mixing is allowed for the subrange $-0.8 < n < 0$.

The energy dependence of the mixing scale will be directly probed as data taking at SuperKamiokande continues,
however several independent tests will be provided by experiments presently under construction.
The large $\nu_e$ mixing allowed in these exotic scenarios 
will be probed by future long baseline, reactor neutrino experiments such as KamLAND \cite{KamLAND},
and also by solar neutrino experiments \cite{newsolar}.  
Long baseline accelerator neutrino experiments running from Fermilab to Soudan (MINOS \cite{MINOS}),
and possibly also Neutrinos to  Gran Sasso (NGS \cite{NGS}) from CERN, 
should be able to fully test these exotic scenarios.

\acknowledgments

We would like to thank L.~de Barbaro and D.~Naples from the CCFR collaboration for useful discussions.  
The work of J.P. is supported by Research Corporation, T.K. is supported by the DOE, 
Grant no.~DE-FG02-91ER40681, and S.M. is supported by the Purdue Research Foundation.

\raggedbottom
\newpage

Table (1).  
The two mixing angles, \(\psi\) and \(\phi\), 
range between 0 and \(\pi/2\).  When one of these mixing
angle is at the limit of its range, 
this three-flavor notation (Eq. (\ref{U})) reduces to
a two-flavor approximation. 
The parameter limits and corresponding equivalent 
two-flavor approximation are given below.
\\
 
\begin{tabular}{| c | c |} \hline
Angle limit  & Equivalent two-flavor mixing \\ \hline
\(\sin^2 \psi = 1.0\)  & \(\nu_e \leftrightarrow \nu_\mu\)   \\
\(\sin^2 \psi = 0.0\)  & \(\nu_e \leftrightarrow \nu_\tau\)   \\
\(\sin^2 \phi = 0.0\) & \(\nu_\mu \leftrightarrow \nu_\tau\)  \\
\(\sin^2 \phi = 1.0\) & no oscillations \\
\hline
\end{tabular}

\raggedbottom
\newpage

Table (2).
Constraints ($90\%$ CL) from the CCFR accelerator experiment \cite{CCFR}
on two-flavor, VEP/LIV oscillation parameters.
\\

\begin{tabular}{| c | c | c |} \hline
Oscillation channel  & Large scale mixing     & Maximal mixing   \\ 
               &  limit on $\sin^2 2 \theta$ &  limit on $|\phi| \Delta \gamma$ \\ \hline
$\nu_\mu \leftrightarrow \nu_e$  &  $2 \times 10^{-3}$ & $2.0 \times 10^{-23}$  \\
$\nu_e \leftrightarrow \nu_\tau$  &  $0.2$ & $3.0 \times 10^{-22}$  \\
$\nu_\mu \leftrightarrow \nu_\tau$  &  $1 \times 10^{-2}$ & $5.0 \times 10^{-23}$  \\
\hline
\end{tabular}

\raggedbottom
\newpage

\begin{figure}

\caption{Bounds on VEP/LIV parameters from the recent CCFR accelerator 
experiment.  The region to the right of the contours is excluded 
for the indicated two-flavor oscillations at 90\% C.L.} 
\label{fig;acc}
\end{figure}

\begin{figure}
\caption{VEP/LIV parameter constraints from atmospheric and accelerator neutrinos.
The vertical axis shows the mixing scale and the horizontal axis shows the $\nu_e$ mixing, 
the $\nu_\mu - \nu_\tau$ neutrino mixing parameter is chosen to minimize the $\chi^2$ value.
Inside the solid contour is the region allowed by the atmospheric contained/partially contained data at 95\% C.L., 
and inside the dashed contour is region allowed by the thrugoing muon data at 95\% C.L.
The best fit point to both data sets simultaneously is indicated by the black dot.
Above the dotted contour is the region excluded by the CCFR accelerator data at 90\% C.L.
There is no set of VEP/LIV parameters which are compatible with all the experimental results.
}
\label{fig;atmos1}
\end{figure}

\begin{figure}
\caption{Constraints on the neutrino oscillation energy exponent from 
atmospheric, accelerator and reactor data.  The chi-squared fit to the SuperKamiokande
contained and thrugoing data is shown for two-flavor $\nu_\mu - \nu_\tau$ mixing (dashed curve)
and for three-flavor mixing (dotted curve).  The mixing scale and relevant mixing angles are
chosen to minimize the chi-squared value.
Above the solid horizontal line is excluded at 99\% C.L.
Also shown are the constraints from reactor and accelerator neutrino experiments.
}
\label{fig;atmos2}
\end{figure}

\begin{figure}
\caption{Parameter region allowed by solar neutrino observations at 90\% CL (solid curve).
Also shown is the region excluded by a three-neutrino analysis of the CCFR accelerator 
experiment at 90\% CL (dotted curve).
The accelerator data exclude the VEP/LIV explanation of the solar neutrino observations.}
\label{fig;solar}
\end{figure}

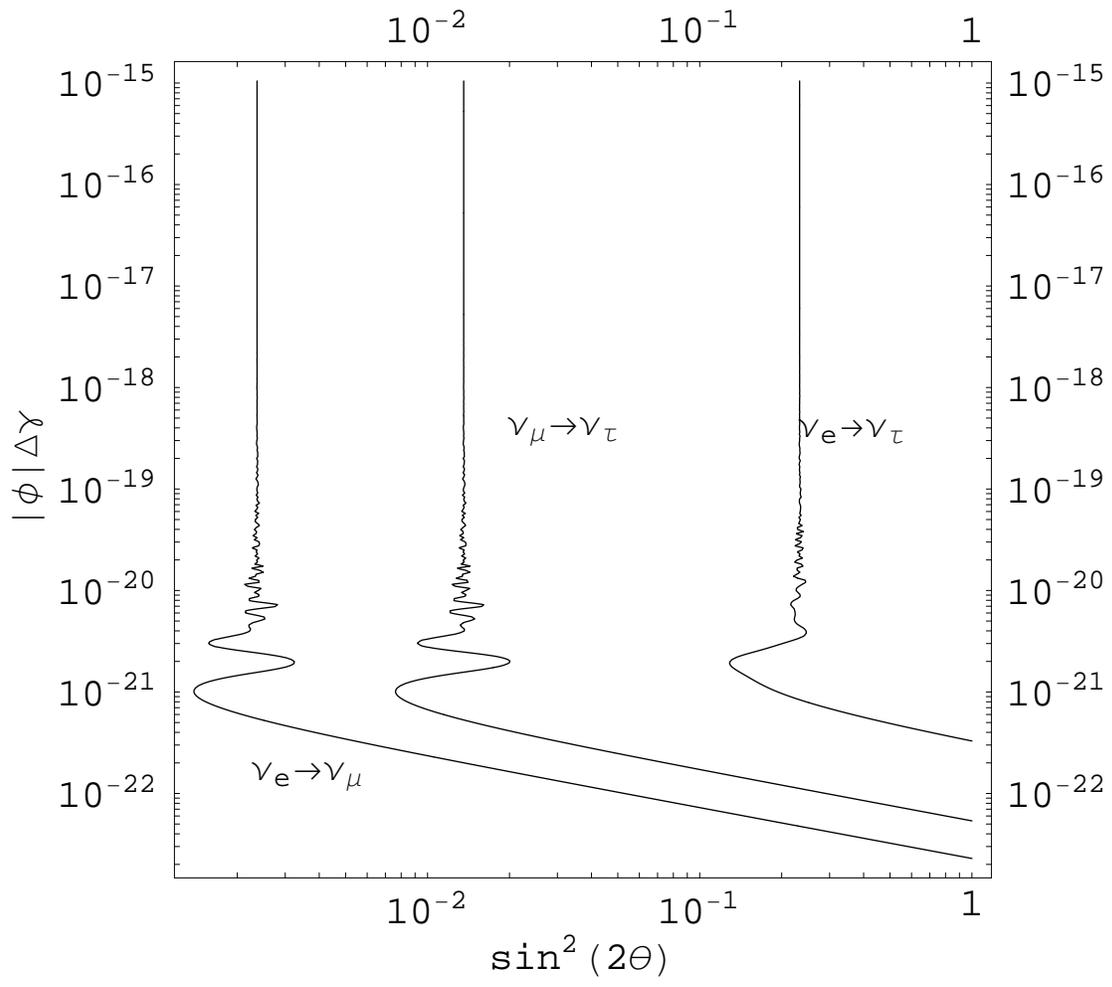
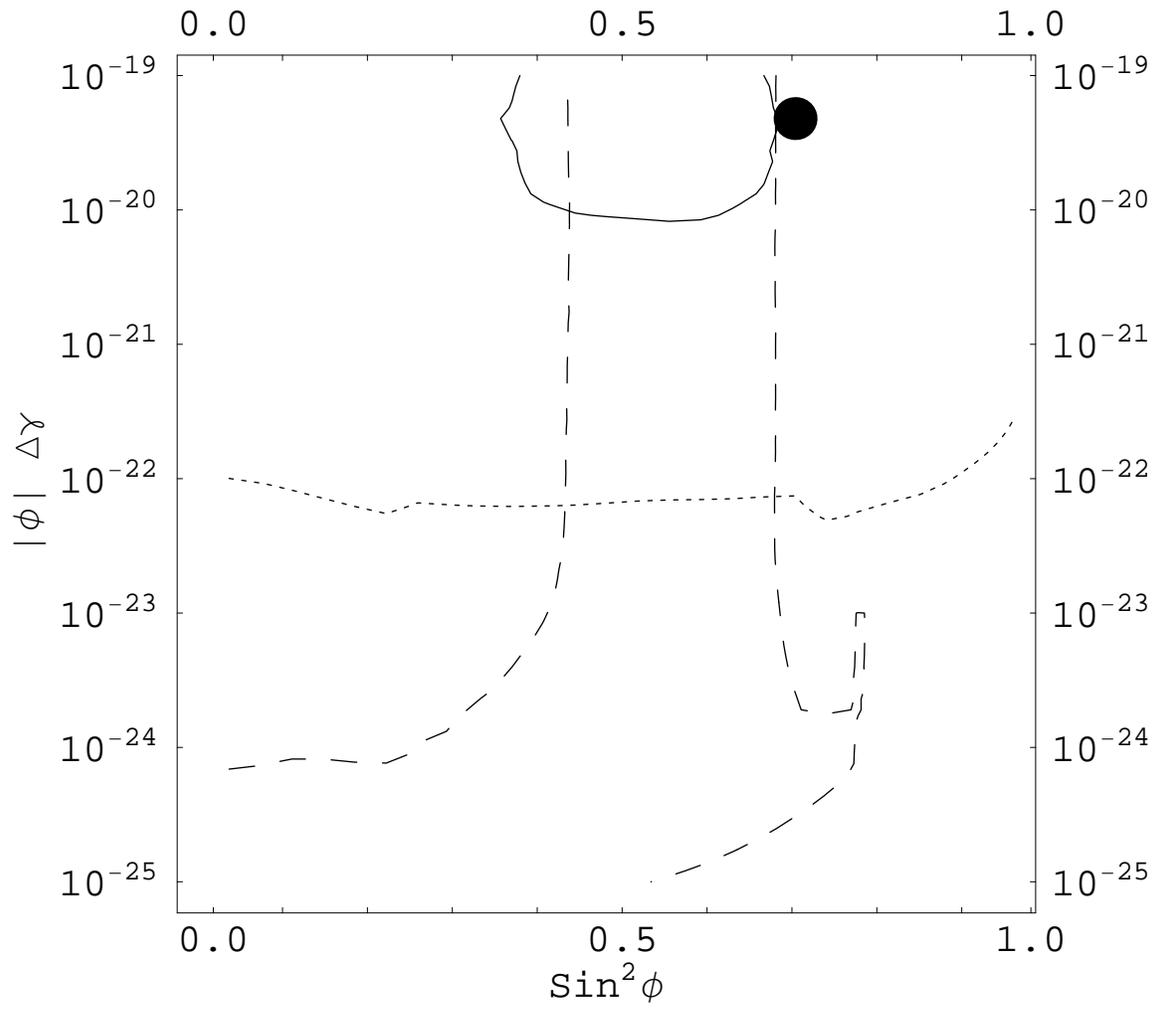
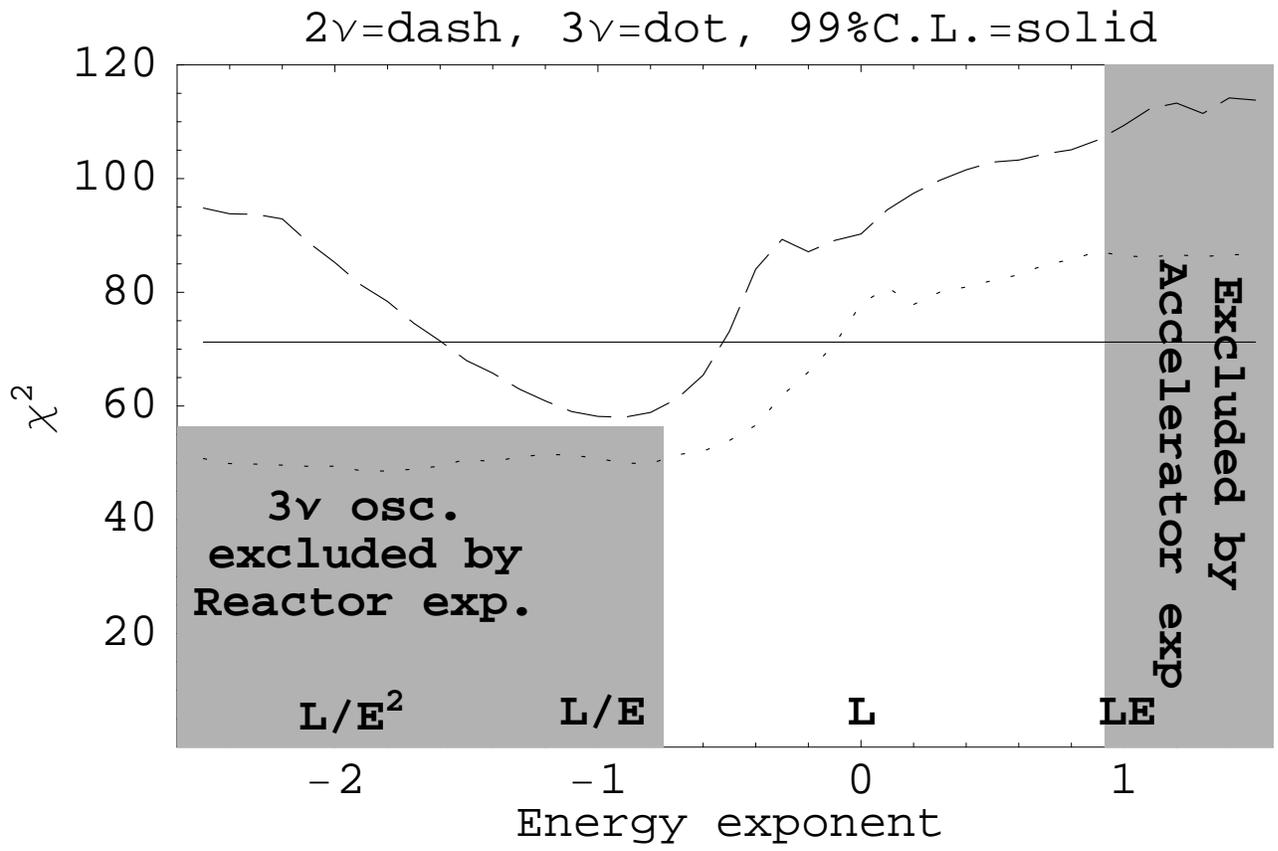
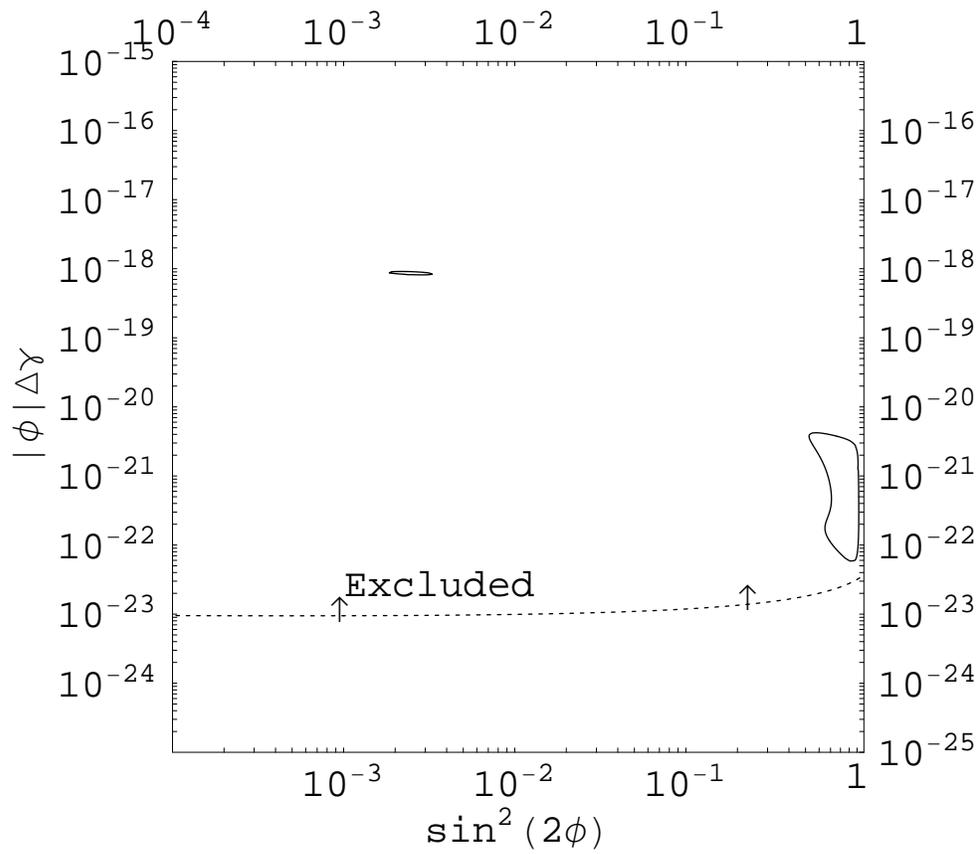
\end{document}